\documentclass[a4paper]{VisionStyle}
\usepackage{epsfig}

\begin{document}

\title{On the nature of X--ray Bright Optically Normal 
galaxies}

\author{A.\,Comastri\inst{1}, M. Brusa\inst{2}, P. Ciliegi\inst{1}, 
M. Mignoli\inst{1}, F. Fiore\inst{3}, R. Maiolino\inst{4}, 
P. Severgnini\inst{5}, A. Baldi\inst{6}, S. Molendi\inst{6},
C. Vignali\inst{7}, F. La Franca\inst{8}, G. Matt\inst{8}, 
G.C. Perola\inst{8} }

\institute{
INAF -- Osservatorio Astronomico di Bologna  via Ranzani 1, I--40127, 
Bologna, Italy
\and 
  Dipartimento di Astronomia Universita' di Bologna, via Ranzani 1, I--40127, 
Bologna, Italy
\and
INAF -- Osservatorio Astronomico di Roma  via Frascati 33, I--00040, 
Monteporzio Catone, Italy
\and
INAF -- Osservatorio Astrofisico di Arcetri  Largo E. Fermi 5
I--50125, Firenze, Italy
\and
INAF -- Osservatorio Astronomico di Brera, via Brera 28, I--20121, Milano,
Italy
\and 
Istituto di Fisica Cosmica -- CNR, via Bassini 15, I--20133, Milano, Italy
\and
Dept. Astron. \& Astroph., Penn State University, 525 Davey Lab., 
State College, PA 16802, USA
\and
Dipartimento di Fisica, Universita' di Roma Tre, via Della Vasca Navale 84, 
I--00146, Roma, Italy  
}

\maketitle 

\begin{abstract}

Multiwavelength observations of the hard X--ray selected sources 
by {\it Chandra} and XMM--{\it Newton} surveys have significantly 
improved our knowledge of the objects responsible of the 
hard X--ray background. 
A surprising finding is the discovery of a population 
of optically dull, X--ray  bright galaxies emerging at 
2--10 keV fluxes of the order of $10^{-14}$ erg cm$^{-2}$ s$^{-1}$.
We present preliminary results of multiwavelength  
observations of a few objects serendipitously discovered
in the field of XMM--{\it Newton} and {\it Chandra} observations 
and intensively studied at longer wavelengths.

\end{abstract}

\section{Introduction}

Thanks to their revolutionary capabilities (arcsec imaging 
and high energy throughput)
{\it Chandra} and XMM--{\it Newton} have opened up a new era in the study 
of the hard X--ray sky. 
Deep Chandra surveys (Brandt et al. 2001, Rosati et al. 2002)
have reached extremely faint fluxes in the 0.5--2 keV and 2--7 keV bands
virtually resolving the entire XRB flux at these energies;
relatively deep XMM--{\it Newton} 
exposures (Hasinger et al. 2001, Baldi et al. 2002) 
have extended by a factor 50 the sensistivity in the 2--10 and 5--10 
keV bands with respect to previous {\tt ASCA} and {\tt BeppoSAX} observations.
Extensive programs of optical identifications showed extremely 
interesting and unexpected results.
The most surprising finding is the discovery of a 
a sizeable number of relatively bright X--ray sources 
spectroscopically identified with early--type ``normal'' galaxies 
without any obvious signature of nuclear activity in the optical spectra
(Fiore et al. 2000; 
Hornschemeier et al. 2001; Giacconi et al. 2001; Barger et al. 2001a,b).
The large X--ray to optical flux ratio, which exceeds by more than 
one order of magnitude the average value of early--type galaxies 
of similar optical luminosity (Fabbiano, Kim, \& Trinchieri 1992), 
and the hard X--ray spectra, determined from the analysis of X--ray colors,  
both suggest that (obscured) AGN activity is taking place in their nuclei.
In principle X--ray spectroscopy could provide 
a stringent test on their nature.  
Unfortunately, all the sources are detected 
with a number of photons which is too small
to apply conventional X--ray spectral fitting techniques and 
to constrain the absorbing column density.
The lack of optical emission lines could be also explained 
if the nuclear light is overshined by either the stellar continuum or 
a non--thermal component, or if they are not efficiently produced. 
A much better understanding of the sources powering X--ray bright, 
optically quiet galaxies can be achieved only by means of multiwavelength
observations.

The first detailed broad band (from radio to X--rays) 
study of what can be considered
the archetypal of this class  objects, the {\it Chandra} source
CXOUJ 031238.9--765134 also known as {\tt P3} (Fiore et al. 2000)
was recently completed by our group (Comastri et al. 2002).
The overall energy distribution is consistent with that of a completely 
hidden possibly Compton thick 
($N_H > $ 1.5 $\times$ 10$^{24}$ cm$^{-2}$)
AGN (Fig. 1). If this is the case most of the 
observed X--ray emission would be due to a scattered/reprocessed 
nuclear component and a luminous ($L_X \simeq$ 10$^{44}$ erg s$^{-1}$) 
AGN should be present at higher energies.
It is plausible that a sizeable population of completely 
hidden AGN could be hosted by optically ``normal'' 
galaxies. Indeed there are claims of several of these sources
in the literature (Mushotzky et al. 2000; Hornschemeier et al. 2001; 
Barger et al. 2001a,b) which seem to support previous findings
based on ROSAT data (Griffiths et al. 1995). 
A better knowledge of their space density and luminosity distribution 
would be extremely important to  trace the 
cosmological evolution of obscured accretion responsible
of a large fraction of the hard X--ray background.

In order to investigate this issue we have collected multiwavelength 
data of a well defined sample of X--ray bright optically ``normal'' galaxies
(hereinafter {\tt XBONG}) selected in the hard ($>$2 keV) X--ray band from
{\it Chandra} and XMM--{\it Newton} observations.
Here we present preliminary results obtained from the analysis of 
optical and radio observations.

\begin{figure}[ht]
  \begin{center}
    \epsfig{file=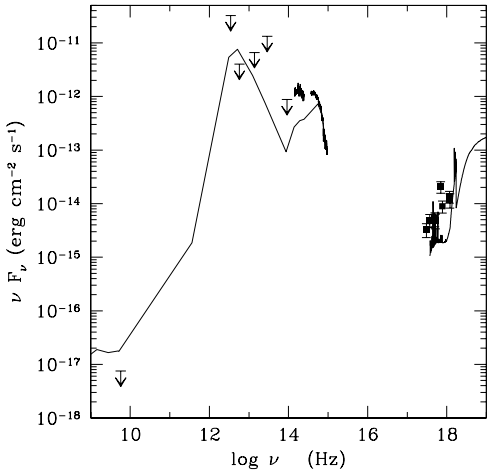, width=7.5cm}
  \end{center}
\caption{The observed SED of P3 including our optical and near--infrared 
spectroscopy, deep radio observations at 5 GHz, the XMM--{\it Newton}
spectrum and IRAS upper limits is compared with that of the highly 
obscured Seyfert 2 galaxy NGC 6240 (Vignati et al. 1999).}
\end{figure}

\section{The XBONG sample}
 
As a starting point for the source selection 
we considered a subsample of the XMM High Energy Large Area Survey:
{\tt HELLAS2XMM} (Baldi et al. 2002) which, at present,
consists of 495 sources detected 
in the hard 2--10 keV band down to a flux limit of 
3 $\times$ 10$^{-15}$ erg cm$^{-2}$ s$^{-1}$ in the field of view 
of 15 XMM--{\it Newton} public observations.
Six of these fields,  containing 147 sources,
were selected for follow--up observations
in the optical band using the ESO 3.6 m and the TNG 3.5 m telescopes.
At the time of writing we have obtained deep R band images 
for all the objects in our sample and radio data at 6 cm 
for one out of the six fields. Spectroscopic identification
are available for 71 sources (Fiore et al., 2002). 
Five of them are classified as {\tt XBONG} on the basis
of a relatively high  X--ray luminosity ($L_{2-10 keV} > 10^{41}$
erg s$^{-1}$) and the lack of 
significant emission lines in their optical spectra.

In order to increase the number of objects 
we have also considered the sample 
of 69 hard X--ray (2--7 keV) selected sources 
compiled by Barger et al. (2001a,b) and based on 
{\it Chandra} observations of three different fields : 
Abell 370 (Bautz et al. 2000), SSA13 (Mushotzky et al. 2000) and 
the CDF--N (Hornschemeier et al. 2001).
Extremely deep radio observations at 20 cm and multicolour optical
imaging data are available for all the sources in this sample while
a spectroscopic redshift has been obtained for 
45 out of the 69 objects.
The optical spectra have been classified by Barger et al. (2001a,b)
into three general categories: (1) broad emission lines, (2) clear signs
of high excitation emission lines and (3) no sign of any of the 
above signatures. There are 10, 15 and 20 sources respectively in each  
of the above mentioned categories.

In order to build an homogeneus and fairly well defined sample of {\tt XBONG} 
we have carefully checked the published optical spectra and applied
the same selection criterium adopted for the {\tt HELLAS2XMM} sample.
Only five out of the 20 galaxies classified by Barger et al. (2001a,b)
in the third category do not show any evidence of line 
emission (with the exception of a weak H$\alpha$ in a few cases).

The sample discussed in the present paper consists of 10
bona--fide {\tt XBONG}. The hard X--ray and optical fluxes are reported 
in Table 1 along with the source redshift and 2--10 keV luminosity. 
Fluxes and luminosities for {\it Chandra} sources have been 
converted from the quoted 2--7 keV energy range  
assuming a power law spectrum with $\Gamma=1.4$. 
R band magnitudes for the  {\it Chandra} selected objects have been 
obtained from the published I band magnitudes assuming 
R--I colors consistent with the values of Fukugita et al. (1995).

\begin{table}[bht]
  \caption{ Multiwavelength properties of the galaxy sample }
  \label{fauthor-E1_tab:tab1}
 \begin{center}
    \leavevmode
    \footnotesize
    \begin{tabular}[h]{lccccc}
      \hline \\[-5pt]
      Source Name & z &  R & B & $F_{2-10 keV}^{a}$ &  log $L_X$ \\[+5pt]
      \hline \\[-5pt]
   XMM0312\#18  & 0.159 & 18.0 & 19.7 & 26.2 & 42.39 \\
   XMM0312\#17  & 0.320 & 17.7 & 19.4 & 28.2 & 43.16 \\
   XMM0312\#8   & 0.050  & 13.7 & 15.4 & 17.8 & 41.07 \\
   XMM2690\#13  & 0.154 & 17.5 & 19.2 & 16.3 & 42.28 \\
   XMM0537\#24  & 0.075 & 21.0 & 22.7 & 45.7 & 41.84 \\
   CXOSSA13\#4  & 0.212 & 20.0 & 22.5 & 15.1 & 42.28 \\
   CXOSSA13\#7  & 0.241 & 19.5 & 21.9 & 7.8 & 42.11 \\
   CXOSSA13\#18  & 0.110 & 16.6 & 18.9 & 3.9 & 41.08 \\
   CXOSSA13\#19  & 0.180 & 17.4 & 19.8 & 3.8 & 41.53 \\
   CXOA370\#10  & 0.360 & 19.3 & 21.9 & 7.4  & 42.41 \\
      \hline \\
      \end{tabular}
\vspace{-0.5cm} 
\flushleft
$^a$: units $10^{-15}$ erg cm$^{-2}$ s$^{-1}$ 
  \end{center}
\end{table}

The X--ray Hubble diagram  for all the spectroscopically identified sources
in the {\tt HELLAS2XMM} and {\it Chandra} samples is reported in Fig.~2.
For the purposes of the present paper 
all the sources with emission lines optical spectra (irrespective
of their intensity and/or width) are reported with the same symbols
and classified as Active Galaxies.
The {\tt XBONG} populate the low redshift, low luminosity end of the 
diagram. Indeed about 75\% of the X--ray sources in the present 
sample with L$_{2-10 keV} < $ 10$^{43}$ erg s$^{-1}$ and z $<$ 0.4 are XBONG.

\begin{figure}[ht]
  \begin{center}
    \epsfig{file=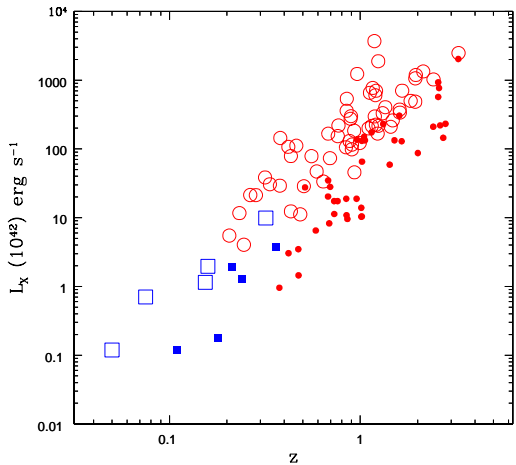, width=7.5cm}
  \end{center}
\caption{The 2--10 keV X--ray luminosity redshift diagram for 
a sample of hard X--ray selected sources from the {\tt HELLAS2XMM}
survey (big open symbols) and {\it Chandra} surveys 
(Barger et al. 2001a,b, small filled symbols). The squares correspond
to {\tt XBONG} while the circles are other classes of emission line objects} 
\end{figure}

This result seems to suggest a sharp transition in the 
optical appearance 
of accreting sources at relatively low luminosities possibly related
to a different mechanism powering the X--ray emission.
Given that low luminosity objects are strongly biased against in
flux limited surveys they might constitute 
an important fraction of the sources responsible of the 
background light.

\section{X--ray and optical properties}

Useful constraints on the nature of the {\tt XBONG} source population 
can be obtained from the analysis of the already available  
optical and X--ray fluxes and from an estimate of their average X--ray
spectral properties inferred from the hardness ratio analysis. 
The R band magnitudes plotted versus the 2--10 keV flux are reported 
in Fig.~3.

\begin{figure}[ht]
  \begin{center}
    \epsfig{file=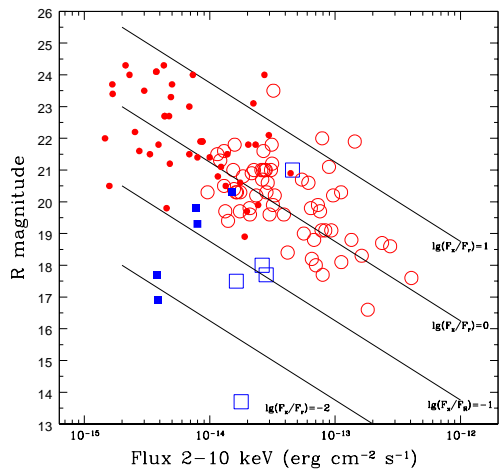, width=7.5cm}
  \end{center}
\caption{The 2--10 keV X--ray flux versus the R magnitude 
for a sample of hard X--ray selected sources from the {\tt HELLAS2XMM}
survey (big open symbols) and {\it Chandra} surveys 
(Barger et al. 2001a,b, small filled symbols). The squares correspond
to {\tt XBONG} while the circles are other classes of AGN. The locii 
of constant $F_X$/F$_{opt}$ are reported with the values as labeled.} 
\end{figure}

The so far identified AGN show a relatively well-defined correlation 
with the optical magnitude around $F_X/F_{opt} \simeq$ 1. This correlation 
is similar to that found by {\tt ROSAT} for soft X--ray selected 
quasars (Hasinger et al. 1998) and extended by {\it Chandra} 
(Alexander et al. 2002) and 
XMM--{\it Newton} observations (Lehmann et al. 2001) also for
hard X--ray selected sources.
The X--ray to optical flux ratio distribution 
of the galaxies in our sample is characterized by a larger 
dispersion and a lower average value 
than that of emission lines AGN suggesting that 
the optical light is dominated by the host galaxy as expected 
if an obscured nucleus is responsible for the observed X--ray emission.
If this were the case, hard X--ray spectra due to strong 
low energy absorption are expected.
The analysis of X--ray colours might give useful information 
in this respect.
The H/S band ratio, where H and S correspond to the fluxes in the 
2--10 keV and 0.5--2  keV energy ranges, has been computed 
for both the {\tt HELLAS2XMM} and the {\it Chandra} samples and reported
in Fig.~4 as a function of the hard X--ray flux.

\begin{figure}[ht]
  \begin{center}
    \epsfig{file=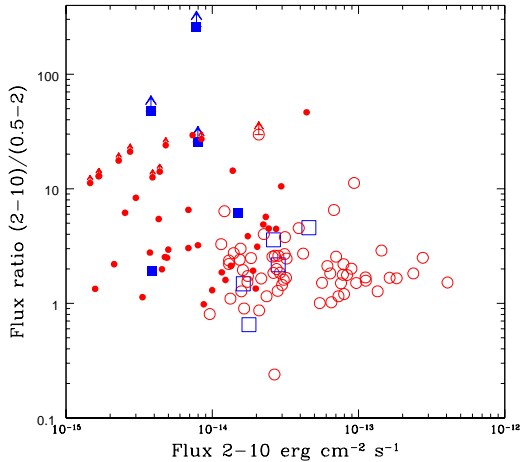, width=7.5cm}
  \end{center}
\caption{The 2--10/0.5--2 keV band ratio is plotted 
versus the X--ray flux. Symbols as in the previous figure.}
\end{figure}

Although three of the faint {\it Chandra} {\tt XBONG} not detected in the
soft X--ray band are characterized by an extremely hard band ratio   
the X--ray spectrum of the brighter galaxies in the 
{\tt HELLAS2XMM} sample is not particulary hard. 
A heavily buried Compton thick AGN may still be present 
if the observed X--ray emission is due to a scattered/reflected 
nuclear component (Comastri et al. 2002).

\par\noindent

Figure 5 shows the relation between the 2--10 keV X--ray luminosity 
and the blue band luminosity (L$_B$) compared with that obtained 
for a nearby sample of early--type galaxies observed with ASCA
(Matsumoto et al. 1997).  The ASCA 0.5--4.5 keV luminosities have been
converted in the 2--10 keV band assuming a power law spectrum
with $\Gamma$ = 1.8 (the average value of the hard component 
in the ASCA sample).
The solid line represents the best fit relation between 
the blue and the hard X--ray luminosities
for bulge--dominated  spiral galaxies (Canizares et al. 1987).
The hard X--ray luminosity of nearby early--type galaxies
is proportional to the optical one and the relation is similar 
to that of spiral galaxies suggesting that high energy X--ray emission
in both early--type and spiral galaxies is due to the integrated 
contribution of low mass X--ray binaries.
These findings have been recently confirmed by high resolution 
{\it Chandra} observations (Loewenstein et al. 2001).

Although a few {\tt XBONG} could be luminous examples of otherwise normal 
elliptical galaxies whose X--ray emission arise from the integrated
contribution of stellar sources such a possibility 
is clearly inconsistent with the most luminous objects in our sample.

\begin{figure}[ht]
  \begin{center}
    \epsfig{file=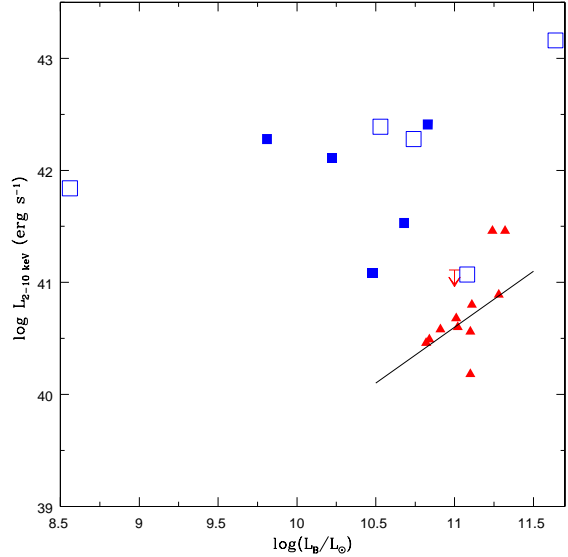, width=8cm}
  \end{center}
\caption{ The 2--10 keV X--ray luminosity is plotted versus the 
optical luminosity in the B band. The filled squares are the
{\it Chandra} selected objects the open squares the {\tt HELLAS2XMM}
 ones. The triangles correspond to the early--type nearby galaxies 
observed by ASCA.} 
\end{figure}

\section{Radio and optical properties}

Deep radio observations at 1.4 GHz are available for all the 5 
{\it Chandra} selected {\tt XBONG}. Although all  
but one have been detected, their radio to optical flux 
ratios qualify these objects as radio quiet (Fig.~6).
Only two of the five {\tt HELLAS2XMM} {\tt XBONG} 
have been observed at 5 GHz with the 
Australia Telescope Compact Array (Brusa et al. 2002).
Interesting enough one object is a relatively bright (S$_{5 GHz}$ = 1.3 mJy)
radio source. The radio and broad band properties of 
the object which was not detected at 5 GHz (P3) have been extensively 
discussed elsewhere (Comastri et al. 2002).

In order to put the present observations on a broader context 
we have collected multiwavelength data for two samples 
of radio selected early--type galaxies from the \linebreak 
ATESP survey (S$_{1.4 GHz} > $ 0.5 mJy;  
Prandoni et al. 2001) and the Marano field radio survey 
(S$_{1.4 GHz} > $ 0.2 mJy; Gruppioni, Mignoli \& Zamorani 1999).
All the objects in these radio selected samples 
are characterized by an almost featureless optical spectrum 
without emission lines and thus are very similar to the 
optical spectra of X--ray selected {\tt XBONG}.

The presence of relatively bright radio sources hosted 
by optically normal galaxies has been pointed out already several 
years ago (Sadler, Jenkins, \& Kotany 1989).
As long as the radio properties are concerned these objects 
are classified as AGN, however the lack of an AGN signature
in the optical band has never been fully understood.

The optical and radio properties of both the  X--ray selected {\tt XBONG} 
and the radio selected galaxies brighter than R $\approx$ 20 
are reported in Fig.~6. 
There are no X--ray observations available for the ATESP 
galaxies while deep ROSAT PSPC observations are
available for the Marano field (Zamorani et al. 1999).
The X--ray selected galaxies are on average weaker in the radio 
band. We note that among the objects with an X--ray detection 
the two brightest radio sources (one from the {\tt HELLAS2XMM} 
sample and the other one in the Marano field) have an optical
to radio flux ratio typical of radio loud AGN.

\begin{figure*}
  \begin{center}
    \epsfig{file=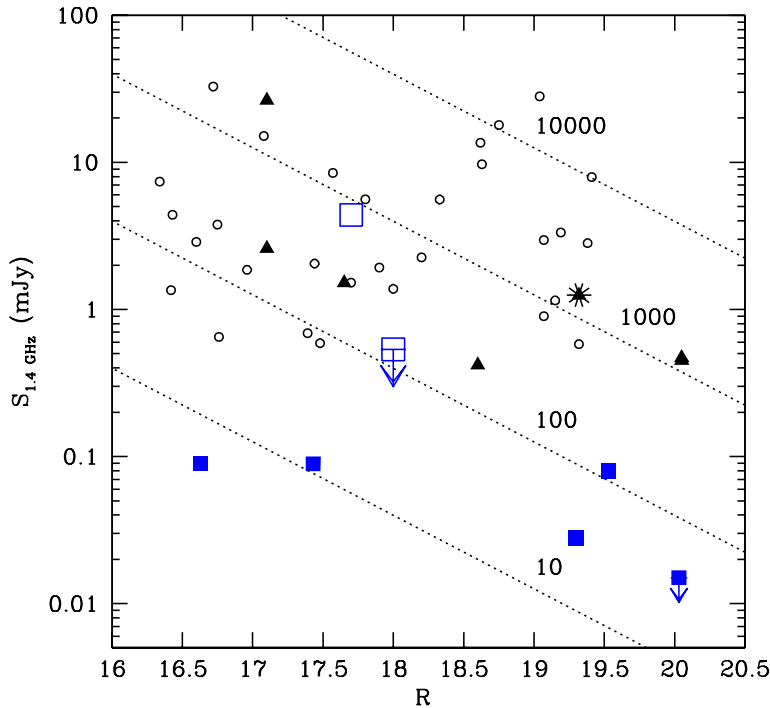, width=12cm}
  \end{center}
\caption{The R band optical magnitude is plotted against the 
20 cm radio flux for the 7 {\tt XBONG} observed in the radio band.
The dotted lines correspond to constant radio to optical flux ratios. 
The filled squares represents the {\it Chandra} selected objects,
the open squares the {\tt HELLAS2XMM} ones. Fluxes at 5 GHz have been 
converted to 1.4 GHz assuming $\alpha$=0.5 ($F_{\nu} \propto \nu^{-\alpha}$).
Small open circles are 
early--type galaxies in the ATESP survey while filled triangles 
represent the galaxies in the Marano field. The star marks the 
object in the latter survey detected in the X--ray band by ROSAT.}
\end{figure*}

\section{Conclusions}

The high X--ray luminosities and the relative ratios between 
X--ray, optical and radio fluxes strongly suggest that 
nuclear activity is going on in almost all the {\tt XBONG}.
The very nature of the mechanism responsible of the observed 
properties is still puzzling.
Based on a detailed multiwavelength study of what can be considered
the prototype of this class of objects (P3; Comastri et al. 2002) 
it has been concluded that an heavily obscured (possibly Compton 
thick) AGN is the most likely explanation.
It is important to point out that such a possibility is not unique
and alternative solutions could still be viable.
Indeed the presence of an advection 
dominated accretion flow (ADAF) or a BL Lac object
cannot be ruled out among the {\tt XBONG} 
which are also relatively luminous at radio frequencies.

Whatever is the nature of the central engine powering the {\tt XBONG} 
X--ray emission, a better knowledge of their luminosity 
and redshift distribution would have important 
consequences for the synthesis of the X--ray background 
and in turn for a better understanding of the 
evolution of accretion powered sources.  
In this context it is interesting to note that both 
Compton thick AGN (Fabian 1999; Wilman, Fabian \& Nulsen 2001)
and ADAF (Di Matteo \& Allen 1999)
have been proposed as an important class of contributors
to the X--ray background.

Even though the bulk of the high energy background is 
produced by a mixture of type 1 and type 2 AGN as predicted 
by ``standard'' population synthesis models 
(Comastri et al. 2001; Gilli Salvati \& Hasinger 2001), 
these models are far from being unique as they 
contain a number of naive assumptions concerning 
the source spectral properties and their cosmological evolution.
The recent {\it Chandra} and XMM--{\it Newton} surveys have already 
suggested that some revisions are needed in order to explain the 
observed redshift distribution which peaks at much lower 
redshifts ($z < $ 1) than the model prediction (Hasinger 2002).

Our knowledge of the {\tt XBONG} population is so far 
limited to the local Universe. 
If a significant fraction of unidentified hard X--ray sources
discovered in deep surveys \linebreak (Alexander et al. 2001) is made of
high redshift {\tt XBONG} (which at $z > 1$ would be extremely faint 
in the otpical band),  
this could alleviate the discrepancy between the observed 
and the predicted redshift distribution.

Sensitive broad band observations of a larger sample of 
{\tt XBONG} are clearly required to settle these issues.

\begin{acknowledgements}

This research has been partially supported by ASI contracts
I/R/103/00 and I/R/107/00, and by the MURST grant Cofin-00--02--36. 
CV also thanks the NASA LTSA grant NAG5--8107 for financial support.  
It is a pleasure to thank Roberto Gilli and Paolo Tozzi for many 
enlightening discussions.  

\end{acknowledgements}


\begin{thebibliography}{}


\bibitem[]{} Alexander, D.M., Brandt, W.N., Hornschemeier, A.E., Garmire, 
G.P., Schneider, D.P., Bauer, F.E., \& Griffiths, R.E., 2001, AJ 122, 2156

\bibitem[]{} Alexander, D.M., Bauer, F.E., Brandt, W.N., Hornschemeier, A.E., 
Vignali, C., Garmire, G.P., \& Schneider, D.P., 2002, these proceedings 
(astro--ph/0202044) 

\bibitem[]{} Baldi, A., Molendi, S., Comastri, A., Fiore, F., Matt, G., 
\& Vignali, C.  2002, ApJ 564, 190

\bibitem []{} Barger, A.J., Cowie, L.L., Mushotzky, R.F., 
\& Richards, E.A., 2001a, AJ 121, 662

\bibitem []{} Barger, A.J., Cowie, L.L., Bautz, M.W., Brandt, W.N., Garmire, 
G.P., Hornschemeier, A.E., Ivison, R.J., \& Owen, F.N. 2001b, AJ 122, 2177

\bibitem []{} Bautz, M.W., Malm, M. R., Baganoff, F. K., Ricker, G. R.,
 Canizares, C. R., Brandt, W. N., Hornschemeier, A. E., \& Garmire, G. P.,
 2000, ApJ 543, L119


\bibitem[]{} Brandt, W.N., Alexander, D.M., Hornschemeier, A.E., et al. 2001, 
AJ 122, 2810 

\bibitem[]{} Brusa, M., Comastri, A., Ciliegi, P., et al. 2002, in preparation

\bibitem[]{} Canizares, C.R., Fabbiano, G., \& Trinchieri, G., 1987, 
ApJ 312, 503

\bibitem[]{} Comastri, A., Fiore, F., Vignali, C., Matt, G., 
Perola, G.C., \& La Franca, F.,  2001, MNRAS 327, 781 

\bibitem[]{} Comastri, A., Mignoli, M., Ciliegi, P., et al., 2002, ApJ in press
(astro--ph/0202080)

\bibitem[]{} Di Matteo, T., \& Allen, S.W., 1999, ApJ 527, L21

\bibitem[]{} Fabbiano, G., Kim, D.W., \& Trinchieri, G. 1992, ApJS 80, 531

\bibitem[]{} Fabian, A.C., 1999, MNRAS 308, L39 

\bibitem[]{} Fiore, F., et al. 2000, New Astronomy 5, 143 (F00)

\bibitem[]{} Fiore, F., et al. 2002, in preparation

\bibitem[]{} Fukugita, M., Shimasaku, K., Ichikawa, T., 1995, PASP 107, 945

\bibitem[]{} Giacconi, R., et al. 2001, ApJ 551, 664  

\bibitem[]{} Gilli, R., Salvati, M., \& Hasinger, G., 2001, A\&A 366, 407 

\bibitem[]{} Griffiths, R.E., Georgantopoulos, I., Boyle, B.J., Stewart, 
G.C., Shanks, T., \& Della Ceca, R., 1995, MNRAS 275, 77 

\bibitem[]{} Gruppioni, C., Mignoli, M., \& Zamorani, G., 1999, MNRAS 304, 199 

\bibitem[]{} Hasinger, G., et al., 1998, A\&A 329, 482

\bibitem[]{} Hasinger, G., et al., 2001, A\&A 365, L45

\bibitem[]{} Hasinger, G., 2002, these proceedings (astro-ph/0202430)

\bibitem[]{} Hornschemeier, A., et al. 2001, ApJ 554, 742 

\bibitem[]{} Loewenstein, M., Mushotzky, R.F., Angelini, L., Arnaud, K.A., 
\& Quataert, E., 2001, ApJ 555, L21 

\bibitem[]{} Matsumoto, H., Koyama, K., Awaki, H., Tsuru, T., Loewenstein, M., 
\& Matsushita, K., 1997, ApJ 482, 133

\bibitem[]{} Mushotzky, R.F., Cowie, L.L., Barger, A.J., \& Arnaud, K.A., 
2000, Nature 404, 459 

\bibitem[]{} Prandoni, I., Gregorini, L., Parma, P., de Ruiter, H.R., 
Vettolani, G., Zanichelli, A., Wieringa, M.H., \& Ekers, R.D., 2001, 
A\&A 369, 787

\bibitem[]{} Rosati, P., et al., 2002, ApJ 566, 667

\bibitem[]{} Sadler, E.M., Jenkins, C.R., \& Kotanyi, C.G., 1989 MNRAS 240, 
591 

\bibitem[]{} Vignati, P., et al. 1999, A\&A 349, L57 

\bibitem[]{} Wilman, R.J., Fabian, A.C., \& Nulsen, P.E.J., 2000, 
MNRAS 319, 583  

\bibitem[]{} Zamorani, G., Mignoli, M., Hasinger, G., Burg, R., 
Giacconi, R., Schmidt, M., Tr\"umper, J., Ciliegi, P., 
Gruppioni, C., Marano, B., 1999, A\&A 346, 731 

\end{thebibliography}
\end{document}